\documentclass[11pt]{article}
\usepackage{geometry}                
\usepackage{amsmath, amsthm, amssymb}
\usepackage{epsfig}
\usepackage{longtable}
\usepackage{graphicx}
\usepackage{amssymb}
\usepackage{epstopdf}
\usepackage{enumerate}
\usepackage{subfig}
\usepackage{setspace}
\usepackage{undertilde}
\usepackage[russian,english]{babel}

\newtheorem{theorem}{Theorem}[section]
\newtheorem{lemma}[theorem]{Lemma}

\newtheorem{proposition}[theorem]{Proposition}

\theoremstyle{definition}
\newtheorem{definition}[theorem]{Definition}

\theoremstyle{remark}

\newcommand{\1}{\mathbf{1}}

\newcommand{\Tr}{\mathrm{Tr}}
\newcommand{\Pf}{\mathrm{Pf}}

\newcommand{\erf}{\mathrm{erf}}

\newcommand{\qdet}{\mathrm{qdet}}

\newcommand{\diag}{\mathrm{diag}}

\newcommand{\bX}{\mathbf{X}}
\newcommand{\bR}{\mathbf{R}}
\newcommand{\bA}{\mathbf{A}}
\newcommand{\bB}{\mathbf{B}}
\newcommand{\bC}{\mathbf{C}}
\newcommand{\bM}{\mathbf{M}}

\newcommand{\bQ}{\mathbf{Q}}

\newcommand{\bD}{\mathbf{D}}

\newcommand{\bY}{\mathbf{Y}}

\newcommand{\bw}{\mathbf{w}}

\begin{document}

\begin{center}{\Large \textbf{A real quaternion spherical ensemble of random matrices}}\\ \vspace{36pt}{\large Anthony Mays\footnote{a.mays@ms.unimelb.edu.au}}\\ \vspace{18pt}\textit{Department of Mathematics and Statistics\\University of Melbourne}
\end{center}

\vspace{36pt}

\begin{abstract}
One can identify a tripartite classification of random matrix ensembles into geometrical universality classes corresponding to the plane, the sphere and the anti-sphere. The plane is identified with Ginibre-type (iid) matrices and the anti-sphere with truncations of unitary matrices. This paper focusses on an ensemble corresponding to the sphere: matrices of the form $\bY= \bA^{-1} \bB$, where $\bA$ and $\bB$ are independent $N\times N$ matrices with iid standard Gaussian real quaternion entries. By applying techniques similar to those used for the analogous complex and real spherical ensembles, the eigenvalue jpdf and correlation functions are calculated. This completes the exploration of spherical matrices using the traditional Dyson indices $\beta=1,2,4$.

We find that the eigenvalue density (after stereographic projection onto the sphere) has a depletion of eigenvalues along a ring corresponding to the real axis, with reflective symmetry about this ring. However, in the limit of large matrix dimension, this eigenvalue density approaches that of the corresponding complex ensemble, a density which is uniform on the sphere. This result is in keeping with the spherical law (analogous to the circular law for iid matrices), which states that for matrices having the spherical structure $\bY= \bA^{-1} \bB$, where $\bA$ and $\bB$ are independent, iid matrices the (stereographically projected) eigenvalue density tends to uniformity on the sphere.
\end{abstract}

\section{Introduction and main results}

This paper follows on from \cite{Mays2011} where the author described a so-called geometrical triumvirate of random matrix ensembles; that is, classes of matrices that (in some sense) live naturally on the three surfaces of constant curvature: the plane, the sphere and the anti-sphere (this categorization was also used in \cite{Kris2006} in a similar context). Through the circular law \cite{Girk1984, Bai1997, GoetTikh2010, TaoVuKris2010}, we identify the Ginibre ensembles \cite{Gini1965} (and other iid ensembles) with the plane, since in the large $N$ limit the eigenvalue density is uniform in some planar disk and zero outside. In fact, in \cite{TaoVu2012} it is shown that the general $n$-point correlation functions are universal in the class of iid matrices, assuming the first four moments are finite.

We identify the sphere with matrices $\bY= \bA^{-1}\bB$, where $\bA$ and $\bB$ are independent iid matrices. The spherical law (conjectured in \cite{ForrMays2011} and \cite{Roge2010}, and proven in \cite{Bord2010}) establishes that the eigenvalues of these matrices (under stereographic projection) are uniformly distributed on the sphere in the limit of large matrix dimension. However, as yet nothing is known about the higher order correlations for the spherical ensembles beyond the Gaussian case. Lastly, the anti-sphere we may identify with truncated ensembles, since in the two studied cases (truncated unitary \cite{ZyczSomm2000, PetzReff2003, Forr2006} and truncated orthogonal \cite{KhorSommZycz2010}) the eigenvalues are uniformly distributed in a hyperbolic space. There is, to date, no `anti-spherical law'.

\begin{figure}[htp]
\begin{center}
\includegraphics[scale=0.36, trim=20 30 30 64, clip=true]{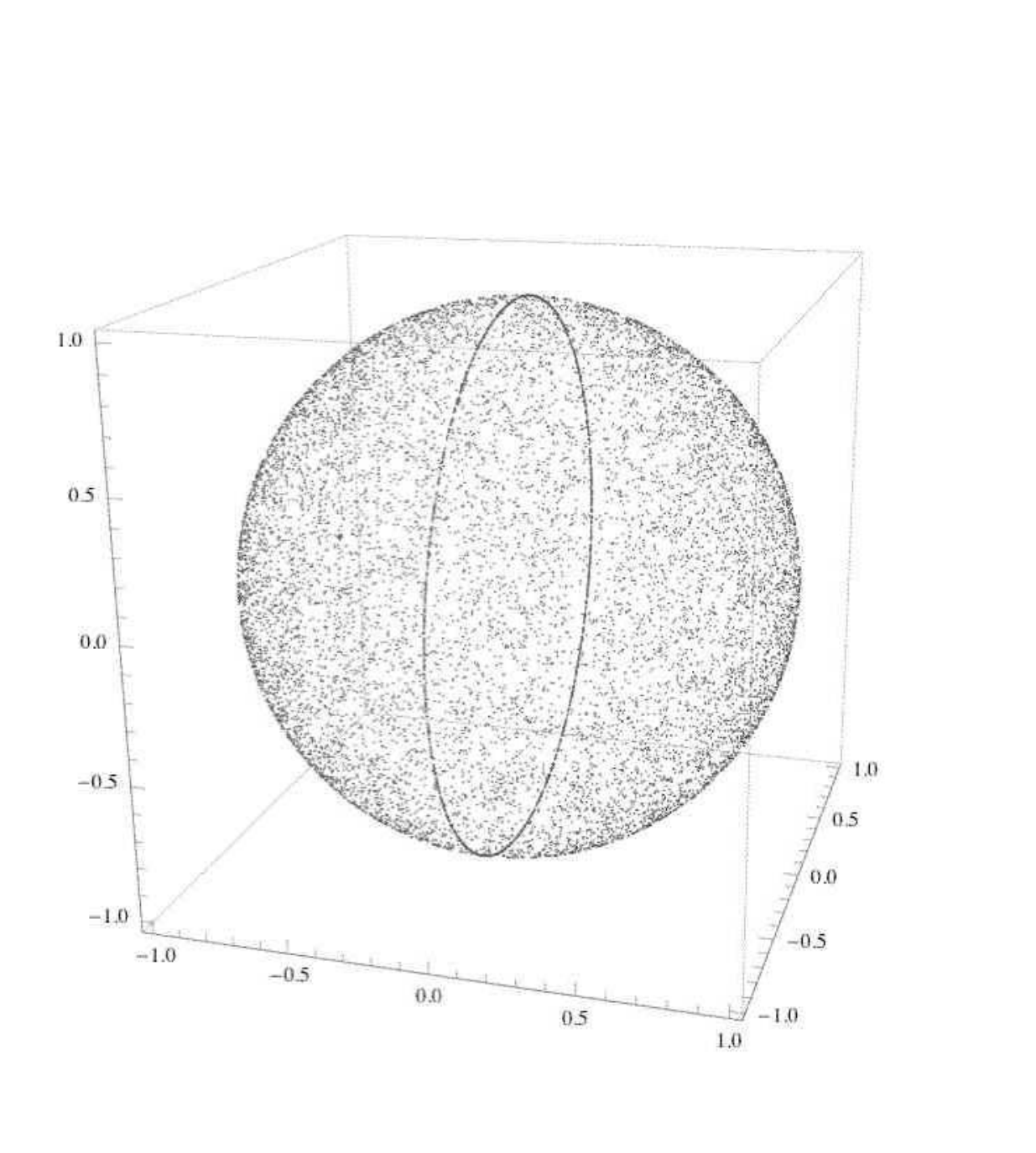}\hspace{45pt}\includegraphics[scale=0.43, trim=50 10 80 54, clip=true]{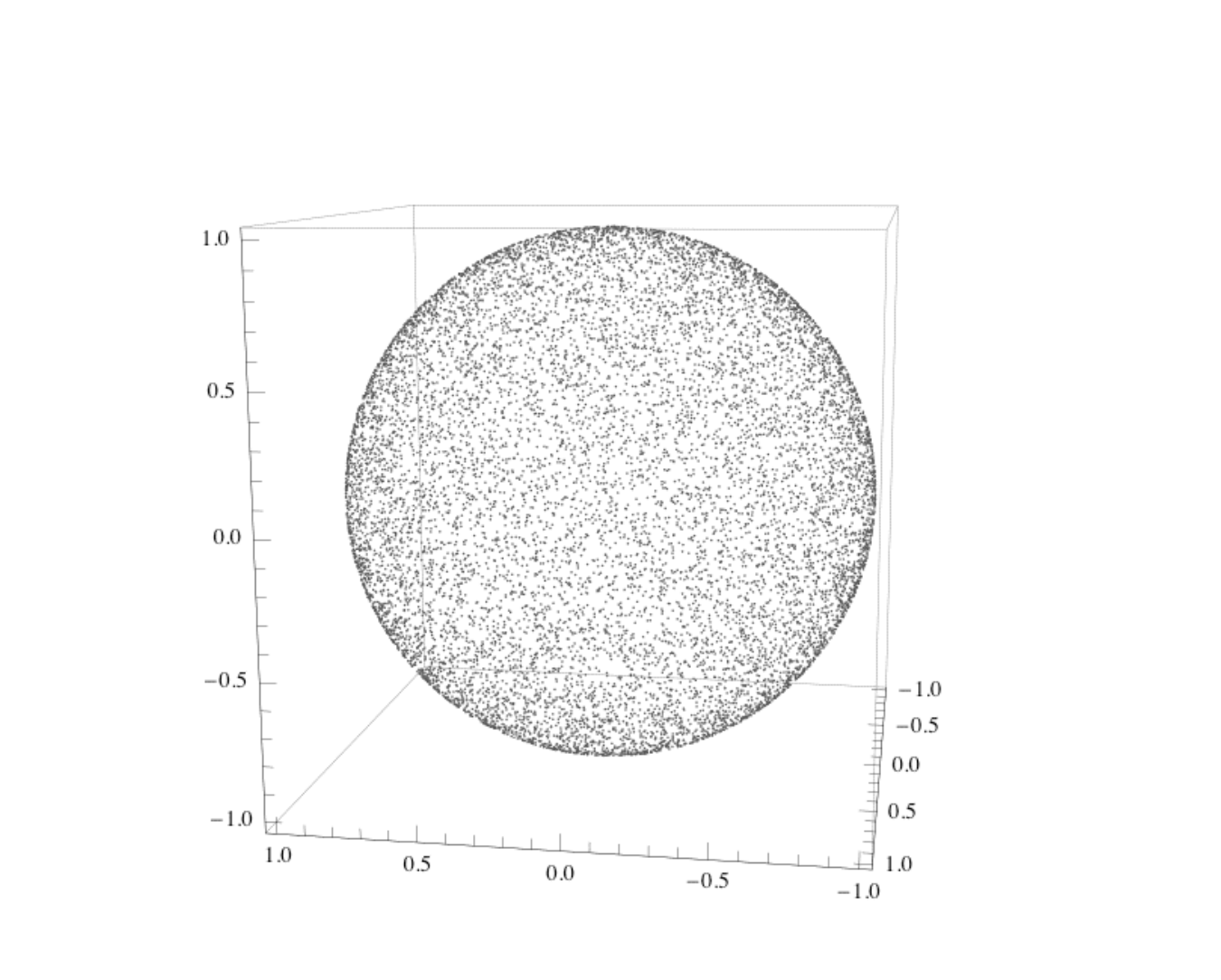}
\end{center}
\caption{Plots of (stereographically projected) eigenvalues for 120 independent $100\times 100$ spherical matrices with real (left), complex (right) entries.}
\label{f:s12figs}
\end{figure}

This paper addresses the outstanding spherical ensemble in the traditional $\beta=1,2,4$ collection: the real quaternion ($\beta=4$) spherical ensemble. Previously, the complex ($\beta=2$) and real ($\beta=1$) cases have been studied in \cite{Fein2004, Kris2006, Kris2009, HKPV2009, ForrMays2011}. To quickly gain some insight into the behaviour of these eigenvalues we can simulate examples (see Figure \ref{f:s12figs}).
Note the ring of eigenvalues in the real spherical ensemble, which is absent in the complex ensemble; this ring corresponds to real eigenvalues, which have a finite probability of appearing in real ensembles. The remaining eigenvalues come in complex conjugate pairs, meaning that the real spherical ensemble has reflective symmetry across the ring (generally the complex eigenvalues of real matrices occur in complex conjugate pairs) and rotational symmetry about the axis through the centre of the ring. The complex ensemble has unbroken rotational symmetry. For the real quaternion spherical ensemble in the present study we will find that there is a depletion of eigenvalues along the ring corresponding to the real axis. Such an effect in the real quaternion Ginibre ensemble follows from the joint eigenvalue probability density function computed in Ginibre's original paper \cite{Gini1965}.

We note that the eigenvalue problem for matrices $\bY=\bA^{-1}\bB$ is equivalent to the generalized eigenvalue problem, that is solutions $\lambda_j$ to
\begin{align}
\nonumber \det(\bB-\lambda\bA)=0,
\end{align}
where $\bA$ is invertible. The real spherical ensemble was discussed in \cite{EdelKostShub1994} in the context of these generalized eigenvalues, where the authors establish that the arctan of the real eigenvalues are uniformly distributed on the unit circle. (See \cite{ForrMays2011} for more discussion on the statistics of the real eigenvalues.)

To approach the real quaternion ensemble, we follow the skew-orthogonal polynomial method used in \cite{ForrMays2011}, which includes an iterated integral technique (see Section \ref{s:ejpdf}) used successfully for the $\beta=2$ spherical ensemble \cite{HKPV2009} (a similar technique was also used in \cite{Gini1965}). For a more detailed description of the techniques used in this paper and the underlying theory, see \cite{Mays2011}.

Since quaternions are crucial to this study, we first provide a quick overview. A quaternion is analogous to a complex number, except that it has four basis elements instead of two. Typically they are written in the form $q=q_0+iq_1+jq_2+kq_3$, with the relations $i^2=j^2=k^2=ijk=-1$, and the $q_l$ are in general complex. We will also use an alternative representation as $2\times 2$ matrices:
\begin{align}
\label{eqn:mat_quat} q=\left[ \begin{array}{cc}
w & x\\
y & z
\end{array}\right],
\end{align}
where $w=q_0+ iq_1, x=q_2+iq_3, y=-q_2+ iq_3, z=q_0- iq_1$. The analogue of complex conjugation for quaternions we denote $q^*=q_0-iq_1-jq_2-kq_3$, or in the matrix representation
\begin{align}
\nonumber q^*=\left[ \begin{array}{cc}
z & -x\\
-y & w
\end{array}\right].
\end{align}
In the case that $q_0, q_1, q_2, q_3\in\mathbb{R}$ we say that $q$ is a \textit{real quaternion} and from (\ref{eqn:mat_quat}), with $\alpha=q_0+ iq_1$ and $\beta=q_2+ iq_3$, we have
\begin{align}
\label{def:real_quats} q=\left[ \begin{array}{cc}
\alpha & \beta\\
-\bar{\beta} & \bar{\alpha}
\end{array}\right],
\end{align}
with conjugate
\begin{align}
\nonumber q^*=\left[ \begin{array}{cc}
\bar{\alpha} & -\beta\\
\bar{\beta} & \alpha
\end{array}\right].
\end{align}
With $\bQ= \left[ q_{j, k} \right]$ we denote by $\bQ^D$ the matrix $[q^*_{k,j}]$, and we call it the \textit{dual} of $\bQ$. If $\bQ=\bQ^D$ then $\bQ$ is said to be \textit{self-dual}. We will regularly use quaternion analogues of the usual matrix trace and determinant \cite{Dyso1970}.
\begin{definition}
\label{d:qqtd}
For an $N\times N$ matrix $\bQ$ with real quaternion entries the \textit{quaternion trace} is defined as the sum of the scalar parts of the diagonal entries
\begin{align}
\label{d:qtr} \mathrm{qTr}\; \bQ := \sum_{j=1}^N (q_0)_{j,j}.
\end{align}
The \textit{quaternion determinant} is defined by
\begin{equation}
\label{def:qdet} \qdet\; \bQ :=\sum_{P\in S_N}(-1)^{N-l}\prod_1^l \; \mathrm{qTr} (q_{ab}q_{bc}\cdot\cdot\cdot q_{sa}),
\end{equation}
where $P$ is any permutation of $(1,...,N)$ that consists of $l$ disjoint cycles of the form $(a\rightarrow b \rightarrow c \rightarrow \cdot\cdot\cdot \rightarrow s \rightarrow a)$.
\end{definition}
Note that the definition (\ref{d:qtr})  gives
\begin{align}
\label{e:qtr} \mathrm{qTr}\; \bQ_{N \times N} = \frac{1}{2} \Tr \; \hat{\bQ}_{2N\times 2N},
\end{align}
where $\hat{\bQ}_{2N\times 2N}$ is the matrix corresponding to $\bQ_{N\times N}$ with the quaternions replaced by their $2\times 2$ representatives (\ref{def:real_quats}). Furthermore, it is shown in \cite{Dyso1970} that with the definition (\ref{def:qdet}) and with $\bQ_{N\times N}$ a self-dual real quaternion matrix
\begin{align}
\label{e:qdet} \qdet\; \bQ_{N\times N} = \big( \det \hat{\bQ}_{2N\times 2N} \big)^{1/2}.
\end{align}
Since we will be mostly using the $2\times 2$ representation for the quaternions we will most often make use of (\ref{e:qtr}) and (\ref{e:qdet}) instead of Definition \ref{d:qqtd}, although we will suppress the `hat' notation for ease of reading.

An element of the real quaternion spherical ensemble is a product of two real quaternion Ginibre matrices. These matrices, as defined in \cite{Gini1965}, are $N\times N$ matrices with independent Gaussian real quaternion entries. For ease of reference, we make an explicit definition.

\begin{definition}
Define standard Gaussian random real quaternion variables as $q=q_0+iq_1+jq_2+kq_3$, where $q_0, q_1, q_2, q_3$ are independent real standard Gaussian variables.

Let $\bA, \bB$ be independent $N\times N$ matrices with standard Gaussian real quaternion entries. Then the \textit{real quaternion spherical ensemble} consists of $N \times N$ matrices of the form $\bY = \bA^{-1}\bB$.

Equivalently, choose $\bA$ and $\bB$ to be independent $2N\times 2N$ matrices containing the $2\times 2$ blocks (\ref{def:real_quats}), where $\alpha, \beta$ are complex numbers with real and imaginary parts being independent standard Gaussian variables. In this case $\bY= \bA^{-1}\bB$ is a $2N \times 2N$ complex matrix.
\end{definition}

In Section \ref{s:mpdf} we calculate the matrix probability density function (pdf) for these matrices. It turns out that we can combine the results in \cite{Fein2004} for the complex and real cases (also later derived independently in \cite{HKPV2009} and \cite{ForrMays2011} respectively) to state the matrix pdfs for the three spherical ensembles (real, complex and real quaternion) in the one form. Note that in \cite{Fein2004} the author shows that a matrix Cauchy distribution is obtained for all spherical matrices, independent of the elemental distribution.
\begin{proposition}
With $\bY$ a real ($\beta=1$), complex ($\beta=2$) or real quaternion ($\beta=4$) spherical matrix, the matrix pdf is
\begin{align}
\label{e:mpdf} \mathcal{P}_{\beta} (\bY) = \pi^{-\beta N^2/2} \prod_{j=0}^{N-1} \frac{\Gamma \left( \frac{N+1+j} {2}\beta\right)} {\Gamma \left( \frac{j+1}{2}\beta \right)} \det (1+\bY \bY^{\dagger})^{-\beta N},
\end{align}
where the `dagger' should be interpreted as `transpose', `Hermitian conjugate' or `quaternion dual' respectively. Note that for $\beta=4$, the determinant is to be understood as a \textbf{quaternion determinant}.
\end{proposition}

\begin{figure}[htp]
\begin{center}
\includegraphics[scale=0.47, trim=40 30 30 54, clip=true]{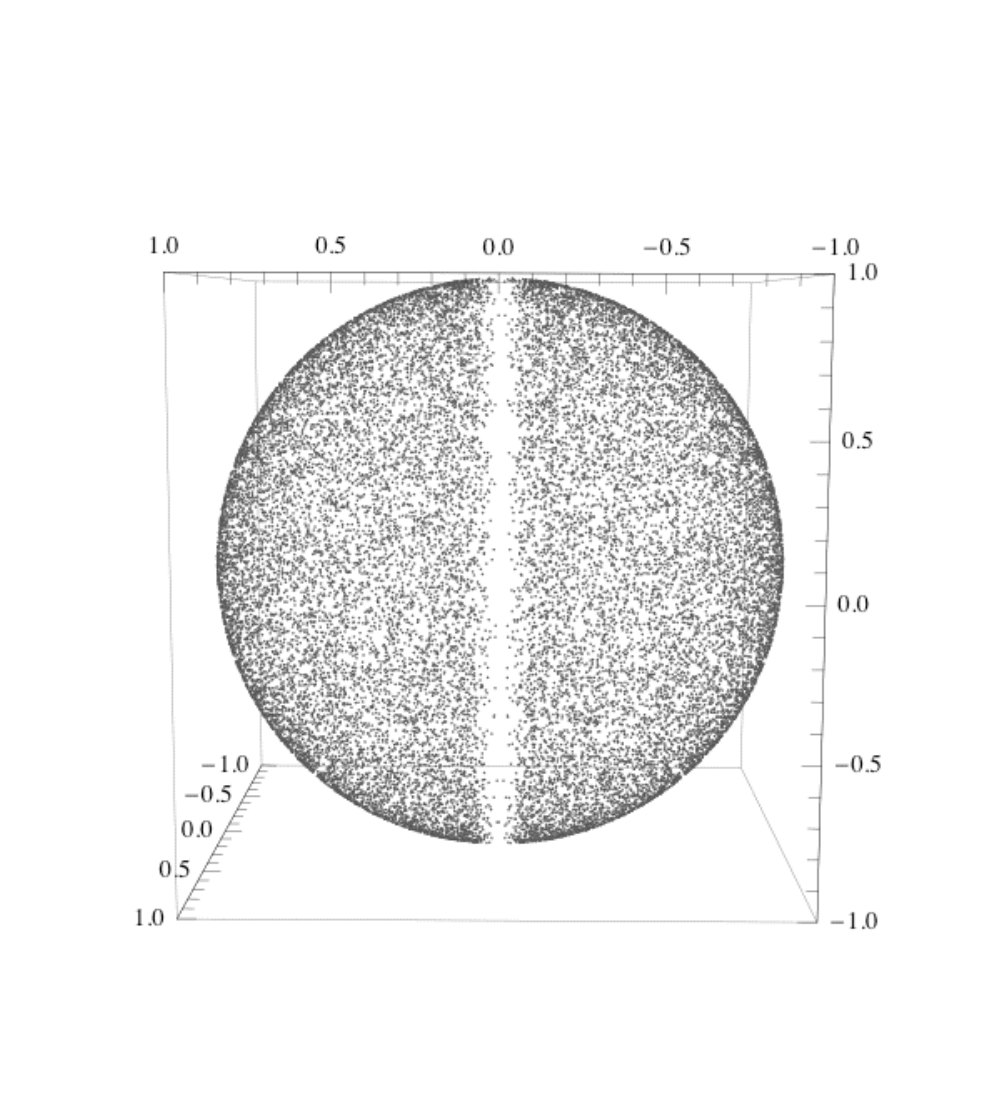}
\end{center}
\caption{(Stereographic) eigenvalue plot for 120 independent $100\times 100$ spherical matrices with real quaternion entries.}
\label{f:s4figs}
\end{figure}

We shall be interested in the eigenvalue distribution; in Figure \ref{f:s4figs} we present a simulation like those in Figure \ref{f:s12figs}, showing that (on stereographic projection) the eigenvalue distribution splits into two symmetrical hemispheres indicating repulsion from the ring corresponding to the real axis, with the complex eigenvalues coming in conjugate pairs (the latter is a general property of self-dual real quaternion matrices).

Using a Schur decomposition, in Section \ref{s:ejpdf} we change variables in (\ref{e:mpdf}) to extract the eigenvalue joint probability density function (jpdf), by applying the iterated integration technique used in \cite{HKPV2009} and \cite{ForrMays2011}. Then, taking advantage of the rotational symmetry about one axis in Figure \ref{f:s4figs}, we apply the fractional linear transformation
\begin{align}
\label{flt} \lambda= \frac{1}{i} \frac{w-1}{w+1},
\end{align}
which takes the upper half plane to the unit disk $\mathbb{D}$, and the lower half-plane to its complement, and we find the eigenvalue jpdf
\begin{align}
\label{transejpdf} \mathcal{Q}(\mathbf{w}) = C_N \prod_{j=1}^N \frac{1}{i \: |w_j|^2} \tau(w_j) \tau \left(\frac{1} {\overline{w}_j} \right) \Delta \left( \mathbf{w}, \mathbf{\frac{1} {\overline{w}}}\right),
\end{align}
with
\begin{align}
\nonumber \mathbf{w}&= \{ w_1, \dots , w_N \}, \qquad w_j\in \mathbb{D},\\
\nonumber \mathbf{\frac{1} {\overline{w}}} &= \left\{ \frac{1} {\overline{w}_1},\dots, \frac{1} {\overline{w}_N} \right\}, \qquad \frac{1} {\overline{w}_j}\in \mathbb{C}\backslash \mathbb{D},\\
\nonumber C_N &= (-1)^{N( N-1)/2} \frac{(\pi)^{-N}}{\Gamma(N+1)} \prod_{j=1}^N {\frac{\Gamma(2N+2)} {(\Gamma(2j))^2}},\\
\nonumber \tau(x) &= \left( \frac{1}{x}\right)^{N-1/2} \frac{\left( \frac{1}{|x|} - |x|\right)^{1/2}} {\left( |x|+\frac{1}{|x|} \right)^{N+1}},\\
\nonumber \Delta (x_1,... ,x_n)&= \prod_{1\leq j<k \leq n} (x_k- x_j).
\end{align}
Note that the factors of $(1/|x|-|x|)$ signify the expected repulsion from the real axis.

The average over this jpdf is calculated in terms of a Pfaffian expression in Section \ref{s:gpf}.
By finding appropriate skew-orthogonal polynomials (Section \ref{s:sops}), and using techniques similar to those in \cite{BoroSinc2009} we calculate the eigenvalue correlation functions in Section \ref{s:correlns}, finding the same structure as in the real quaternion Ginibre ensemble \cite{Gini1965, Kanz2002}. 

In Section \ref{s:dens} we find the eigenvalue density to be
\begin{align}
\nonumber \rho_{(1)}(w)&= \frac{1}{r^2} \frac{\left( \frac{1}{r}- r \right)} {\left( r + \frac{1} {r} \right)^{2N+2} } \frac{(2N+1) (2N)} {\pi} \sum_{j=0}^{N-1} {2N-1 \choose j} \frac{\left( r^{2j+1 -2N} - r^{2N-1-2j} \right)} {2N-2j-1}\\
\label{sxx1} &=\frac{1}{r^2} \frac{\left( \frac{1}{r}- r \right)} {\left( r + \frac{1} {r} \right)^{2N+2} } \frac{2^{2N-1} (2N+1) N}{\pi} \int_{0}^{1- r^2} (1-\gamma)^{-N-1/2} (1-\gamma/2)^{2N-1} d\gamma,
\end{align}
where $w=r\: e^{i\theta}$, which for large $N$ tends to
\begin{align}
\label{e:limdens} \rho_{(1)} (w) \mathop{\sim}\limits_{N\to \infty} \frac{2 N}{\pi (1+r^2)^2},
\end{align}
which is the expected analogue of \cite[Eqn.  (59)]{ForrMays2011}. This limiting density tells us that, on stereographic projection, we have a uniform density of eigenvalues on the sphere in the limit of large matrix dimension. This is a consequence of the spherical law, which was discussed at the beginning of this introduction.

With $\lambda_j\mapsto \lambda_j /\sqrt{2N}$ in (\ref{b4ejpdf}), then we note that (for large $N$) $\mathcal{Q}(\vec{\lambda})$ approaches
\begin{align}
\label{e:limejpdf} \tilde{C}_N \prod_{j=1}^N e^{-|\lambda_j|^2} |\lambda_j -\overline{\lambda}_j|^2 \prod_{1\leq j < k \leq N} |\lambda_k- \lambda_j|^2 |\lambda_k- \overline{\lambda}_j|^2,
\end{align}
where $\tilde{C}_N$ is independent of the $\lambda_j$. This expression (\ref{e:limejpdf}) is the same as the eigenvalue jpdf for the Ginibre ensemble of real quaternion matrices \cite[Eq. (2.28)]{Gini1965}, and so we may expect to recover the same bulk correlation functions (see \cite{Kanz2002}) from our correlation functions here by a suitable scaled limit. In Section \ref{s:slims} we apply the rescaling (\ref{e:slims}) and obtain Proposition \ref{p:slims}, which matches the result for the $\beta=4$ Ginibre case.

\section{Matrix probability density function}
\label{s:mpdf}

With $\bA$ and $\bB$ each being $N\times N$ matrices from the real quaternion Ginibre ensemble (that is, having iid standard normal real quaternion entries) the joint distribution of $\bY=\bA^{-1}\bB$ is
\begin{align}
\label{e:mpdf1} (2\pi)^{-4N^2} e^{- \mathrm{qTr} (\bA\bA^{D} +\bB\bB^{D} )/2} (d\bA) (d\bB) = (2\pi)^{-4N^2} e^{- \mathrm{Tr} (\hat{\bA} \hat{\bA}^{\dagger} +\hat{\bB}\hat{\bB}^{\dagger})/4 } (d\hat{\bA}) (d\hat{\bB}),
\end{align}
where we have changed to the $2\times 2$ representation for the RHS. (For the remainder of this paper we will suppress the `hat' notation, however we will endeavour to preserve the distinction between $D$ for quaternion dual and $\dagger$ for Hermitian conjugate, which, along with `qTr' and `qdet', will serve to indicate which representation we are using.) The plan now is to change variables to $\bY$ and integrate out the remaining degrees of freedom. To achieve this we will follow the real quaternion analogue of the procedure used in \cite{HKPV2009} and \cite{ForrMays2011}. First we list some required lemmata specific to real quaternion matrices.

\begin{lemma}[\cite{Olki2002}]
\label{l:xca}
Let $\bX$ be an $n\times m$ ($n\geq m$) matrix with real quaternion entries, and $\bA= \bX^D \bX$. Then
\begin{align}
\nonumber (d\bX)= \tilde{c} \; (\qdet \; \bA)^{2(n-m)+1} (d\bA),
\end{align}
where $\tilde{c}$ is independent of $\bX$.
\end{lemma}

\begin{lemma}[\cite{Math1997}]
\label{l:adma}
With $\bA$ and $\bM$ both $n\times n$ real quaternion matrices (with $\bA$ non-singular) and $\bX= \bA^D \bM \bA$ we have
\begin{align}
\nonumber (d\bX)= (\qdet \; \bA^D \bA)^{2n-1} (d\bM).
\end{align}
\end{lemma}

\begin{lemma}[\cite{Forr2010}]
\label{e:Selb}
A corollary of the Selberg integral \cite{Selb1944} is the evaluation
\begin{align}
\nonumber &\int_0^{\infty}dx_1\cdot\cdot\cdot \int_0^{\infty}dx_N \prod_{l=1}^N x_l^{2 a} e^{-2 x_l}\prod_{1\leq j < l \leq N}|x_l -x_j|^{4}\\
\nonumber &=2^{-2N( N+ a)} \prod_{j=0}^{N-1} \Gamma \left(3+2j \right) \Gamma (2a +1+ 2j).
\end{align}
\end{lemma}

\begin{lemma}
\label{l:xAy}
Let $\bB_{m\times n}= x_{m\times m}\bA_{m\times n} y_{n\times n}$, where each of the matrices consist of $2\times 2$ real quaternions (\ref{def:real_quats}), then 
\begin{align}
\nonumber (d\bB_{m\times n})= |\qdet \; x_{m\times m} |^{4n} |\qdet \; y_{n\times n} |^{4m} (d\bA_{m\times n}).
\end{align}
\end{lemma}
This last lemma is the real quaternion analogue of the statement about real matrices in Theorem 2.1.5 of \cite{Muir2005}. From the working there we can verify this lemma by noting that $x$ acts on the columns of $\bA$ (of which there are $n$) and $y$ on the rows (of which there are $m$), and then each quaternion has 4 independent components. 

Letting $\bC:= \bA\bA^D$ (which is positive definite) we see from Lemma \ref{l:xca} that
\begin{align}
\label{e:ctilde} (d\bA)= \tilde{c}\; \qdet\; \bC \; (d\bC),
\end{align}
and also, letting $\bB=\bA \bY$, by Lemma \ref{l:xAy} we have $(d\bB) = |\qdet\; \bA|^{4N} (d\bY)$\\ $= (\det \bC)^N (d\bY)$, and so (\ref{e:mpdf1}) becomes
\begin{align}
\label{e:mpdf2} (2\pi)^{-4N^2} e^{-\Tr (\bC (\1+ \bY\bY^{\dagger} ))} (\det \bC)^{N+1/2} (d\bC) (d\bY).
\end{align}
To calculate $\tilde{c}$ we multiply both sides of (\ref{e:ctilde}) by $e^{-(\Tr \bA\bA^{\dagger})/4}$ and integrate over the whole domain
\begin{align}
\nonumber \int e^{-(\Tr \bA \bA^{\dagger})/4} (d\bA) = \tilde{c} \int_{\bC >0} e^{-(\Tr \bC) /4} (\det \bC)^{1/2} (d\bC),
\end{align}
where the positive definiteness of $\bC$ is indicated by $\bC >0$ in the integral terminal. We note that the LHS is just an integral over $4N^2$ independent standard Gaussians and so
\begin{align}
\nonumber \tilde{c}= \frac{(2\pi)^{2N^2}} {\int_{\bC >0} e^{-(\Tr \bC) /4} (\det \bC)^{1/2} (d\bC)}.
\end{align}
Let $\bD:= (\1+ \bY\bY^{\dagger} )^{1/2} \bC (\1+ \bY\bY^{\dagger} )^{1/2}$ in (\ref{e:mpdf2}) and then integrate over $\bD$ to obtain
\begin{align}
\nonumber (2\pi)^{-2N^2} \det (\1+ \bY\bY^{\dagger} )^{-2N} \frac{\int_{\bD >0} e^{-(\Tr \bD) /4} (\det \bD)^{N+1/2} (d\bD)} {\int_{\bC >0} e^{-(\Tr \bC) /4} (\det \bC)^{1/2} (d\bC)} \; (d\bY),
\end{align}
where we used Lemma \ref{l:adma} to obtain the Jacobian for changing variables from $\bC$ to $\bD$. Since both $\bC$ and $\bD$ are self-dual real quaternion matrices we can make use of the well known Jacobians for changing variables from the matrix entries to the matrix eigenvalues (see for example \cite[Chapter 1.3]{Forr2010}) so that the ratio of integrals becomes
\begin{align}
\nonumber \frac{\int_{0}^{\infty}d\lambda_1 \dots \int_{0}^{\infty} d\lambda_N \prod_{j=1}^N \lambda_j^{2N+1} e^{-\lambda_j/2} \prod_{j<k} |\lambda_k- \lambda_j|^4} {\int_{0}^{\infty}d\lambda_1 \dots \int_{0}^{\infty} d\lambda_N \prod_{j=1}^N \lambda_j\; e^{-\lambda_j/2} \prod_{j<k} |\lambda_k- \lambda_j|^4}.
\end{align}
By changing variables $\lambda_j\mapsto 4\lambda_j$ we can apply Lemma \ref{e:Selb} to evaluate the integrals, giving the final expression
\begin{align}
\nonumber \pi^{-2 N^2} \prod_{j=0}^{N-1} \frac{\Gamma \left( 2(N+1+j) \right)} {\Gamma \left( 2(j+1) \right)} \; \qdet (1+\bY_N \bY_N^{D})^{-4 N} (d \bY).
\end{align}
By comparing this to the analogous results in the real and complex cases we can write them in the single equation (\ref{e:mpdf}).

\section{Eigenvalue jpdf}
\label{s:ejpdf}

Here we will change variables in the matrix pdf (\ref{e:mpdf}) from the elements of $\bY$ to the eigenvalues of $\bY$. Using the $2\times 2$ representation of quaternions (\ref{def:real_quats}) apply the Schur decomposition $\bY= \bQ^{\dagger} \bR_{2N} \bQ$ (see \cite{Akem2005} for a detailed description of this in the real quaternion case), with $\bR_{2N}$ an upper triangular real quaternion matrix
\begin{align}
\nonumber \bR_{2N}= \left[\begin{array}{cccc}
\left[ \begin{array}{cc}
0 & \lambda_1\\
\overline{\lambda}_1 & 0
\end{array}\right] & \left[ \begin{array}{cc}
r_{1,2} & s_{1,2}\\
-\overline{s}_{1, 2} & \overline{r}_{1,2}
\end{array}\right] & \dots & \left[ \begin{array}{cc}
r_{1,N} & s_{1,N}\\
-\overline{s}_{1, N} & \overline{r}_{1,N}
\end{array}\right]\\
\left[ \begin{array}{cc}
0 & 0\\
0 & 0
\end{array}\right] & \left[ \begin{array}{cc}
0 & \lambda_2\\
\overline{\lambda}_2 & 0
\end{array}\right] & \dots & \left[ \begin{array}{cc}
r_{2,N} & s_{2,N}\\
-\overline{s}_{2, N} & \overline{r}_{2,N}
\end{array}\right]\\
\vdots & \vdots & \ddots & \vdots\\
\left[ \begin{array}{cc}
0 & 0\\
0 & 0
\end{array}\right] & \left[ \begin{array}{cc}
0 & 0\\
0 & 0
\end{array}\right] & \dots & \left[ \begin{array}{cc}
0 & \lambda_{N}\\
\overline{\lambda}_{N} & 0
\end{array}\right]
\end{array}\right],
\end{align}
where $\lambda_j$ and $\overline{\lambda}_j$ are the $N$ complex conjugate-paired eigenvalues of $\bY$, $r_{j,k}$ and $s_{j,k}$ are complex numbers, and $\bQ\in Sp(N)/(U(1))^N$ is a symplectic (\textit{ie}.~unitary real quaternion) matrix (using the $2\times 2$ representation). Since the decomposition is only unique up to ordering on the eigenvalues, we divide through by $2^N \Gamma[N+1]$. The Jacobian for this change of variables is \cite{Meht2004, Akem2005}
\begin{align}
\nonumber (d\bY) = (\bQ^{\dagger}d\bQ) (d\bR_{2N}) |\tilde{\Delta} (\vec{\lambda})|^2,
\end{align}
where $\tilde{\Delta} (\vec{\lambda}):= \prod_{j=1}^N (\lambda_j- \overline{\lambda}_j) \prod_{1\leq j< k\leq N} (\lambda_j- \lambda_k) (\lambda_j- \overline{\lambda}_k)$ and
$(d\bX)$ represents the wedge product of all the independent elements of a matrix of differentials $d\bX$.
Then
\begin{align}
\label{e:matpdf1} \mathcal{P}_{4}(\bY) (d\bY)= \frac{(\bQ^{\dagger}d\bQ) (d\bR_{2N})}{\pi^{2N^2} 2^N \Gamma(N+1)} \prod_{j=0}^{N-1} \frac{\Gamma \left( 2(N+1+j) \right)} {\Gamma \left( 2(j+1) \right)}\; |\tilde{\Delta} (\vec{\lambda})|^2 \det (\1_{2N}+\bR_{2N} \bR_{2N}^{\dagger})^{-2 N}.
\end{align}

We will now apply the iterated integration technique used in \cite{HKPV2009} and \cite{ForrMays2011} to integrate over the strictly upper triangular components of $\bR_{2N}$. First we separate out the last column
\begin{align}
\nonumber \bR_{2N}=\left[\begin{array}{cc}
\bR_{2N-2} & u_{2N-2}\\
0 & z_N
\end{array}\right],
\end{align}
where $z_N$ is the bottom right $2\times 2$ block on the diagonal, which contains the eigenvalues $\lambda_N, \overline{\lambda}_N$, and $u_{2N-2}$ is of size $(2N-2)\times 2$; explicitly 
\begin{align}
\nonumber u_{2N-2}= \left[ \begin{array}{cc}
r_{1, N-1} & s_{1, N}\\
- \overline{s}_{1,N} & \overline{r}_{1, N}\\
\vdots & \vdots\\
r_{N-1, N-1} & s_{N-1, N}\\
- \overline{s}_{N-1,N} & \overline{r}_{N-1, N}
\end{array}
\right].
\end{align}
Then the determinant in (\ref{e:matpdf1}) becomes
{\footnotesize
\begin{align}
\nonumber &\det (\1_{2N}+\bR_{2N} \bR_{2N}^{\dagger})^{-2 N} = \det  \left(
\left[ \begin{array}{cc}
\1_{2N-2}+ \bR_{2N-2}\bR_{2N-2}^{\dagger} + u_{2N-2} u_{2N-2}^{\dagger} & u_{2N-2} z_N^\dagger\\
z_N u_{2N-2}^{\dagger} & \1_2 + z_N z_N^{\dagger}
\end{array}\right] \right)^{-2N}\\
\nonumber &= \det \left(
\left[ \begin{array}{cc}
\1_{2N-2}+ \bR_{2N-2}\bR_{2N-2}^{\dagger} + u_{2N-2} u_{2N-2}^{\dagger} - u z_N^{\dagger} (1+z_Nz_N^{\dagger})^{-1}z_N u_{2N-2}^{\dagger} & 0\\
z_N u_{2N-2}^{\dagger} & \1_2 + z_N z_N^{\dagger}
\end{array}\right] \right)^{-2N}\\
\nonumber &= \det \left( \1_2+ z_N z_N^{\dagger} \right)^{-2N} \det\left( \1_{2N-2} + \bR_{2N-2}\bR_{2N-2}^{\dagger} + u_{2N-2} u_{2N-2}^{\dagger} - u z_N^{\dagger} (1+z_Nz_N^{\dagger})^{-1}z_N u_{2N-2}^{\dagger} \right).
\end{align}
}We use the identity $z^{\dagger} (1+z z^{\dagger}) z= 1- (1+z^{\dagger} z)^{-1}$ to write the second determinant above as
\begin{align}
\nonumber &\det \left( \1_{2N-2} +\bR_{2N-2}\bR_{2N-2}^{\dagger} + u_{2N-2} (\1_2 +z_N^{\dagger} z_N)^{-1} u_{2N-2}^{\dagger} \right)\\
\nonumber &= \det \left( \1_{2N-2} +\bR_{2N-2}\bR_{2N-2}^{\dagger} \right) \det \left(\1_2 + v_{2N-2}^{\dagger} v_{2N-2} \right),
\end{align}
where
\begin{align}
\label{e:vu} v_{2N-2} := (\1_{2N-2} +\bR_{2N-2} \bR_{2N-2}^{\dagger})^{-1/2}\; u_{2N-2}\; (\1_2 +z_N^{\dagger} z_N)^{-1/2}.
\end{align}
So (\ref{e:matpdf1}) becomes
\begin{align}
\nonumber &\mathcal{P}_{4}(\bY) (d\bY)= \frac{(\bQ^{\dagger}d\bQ) (d\bR_{2N})}{\pi^{2N^2} 2^N \Gamma(N+1)} \prod_{j=0}^{N-1} \frac{\Gamma \left( 2(N+1+j) \right)} {\Gamma \left( 2(j+1) \right)} |\tilde{\Delta} (\vec{\lambda})|^2 \det\left( \1_2+ z_N z_N^{\dagger}\right)^{-2 N} \\
\label{P4split} &\times \det \left( \1_{2N-2} +\bR_{2N-2}\bR_{2N-2}^{\dagger} \right)^{-2 N} \det \left(\1_2 + v_{2N-2}^{\dagger} v_{2N-2} \right)^{-2 N}.
\end{align}
Note that
\begin{align}
\nonumber (d\bR_{2N}) = \prod_{j=1}^N d\lambda_j \bigwedge_{s=1}^{N-1} (du_{2N-2s})
\end{align}
and we want to integrate over the independent elements of $u_{2N-2}$, \textit{ie}.~we want to calculate
\begin{align}
\label{itint1} \int (du_{2N-2}) \det \left(\1_2 + v_{2N-2}^{\dagger} v_{2N-2} \right)^{-2 N}.
\end{align}

Applying Lemma \ref{l:xAy} to (\ref{e:vu}) we have
\begin{align}
\nonumber (du_{2N-2}) &= \qdet (\1_{N-1} +\bR_{N-1} \bR_{N-1}^{D})^2 \qdet (1 +z_N z_N^{D})^{2N-2} (dv_{2N-2})\\
\label{uvjac} &= \det (\1_{2N-2} +\bR_{2N-2} \bR_{2N-2}^{\dagger}) \det (\1_2 +z_N z_N^{\dagger})^{N- 1} (dv_{2N-2}).
\end{align}
For ease of notation, let $a_{2N-2}:= v_{2N-2}^{\dagger} v_{2N-2}$, so
\begin{align}
\nonumber a_{2N-2}= \left[\begin{array}{cc}
\sum_{j=1}^{N-1} |r_{j,N}|^2 +|s_{j,N}|^2 & 0\\
0 & \sum_{j=1}^{N-1} |r_{j,N}|^2 +|s_{j,N}|^2
\end{array} \right].
\end{align}
We can again make use of Lemma \ref{l:xca} by applying it to $a_{2N-2}$ giving
\begin{align}
\label{vacovs} (dv_{2N-2}) = \tilde{c}\; (\qdet \; a_{2N-2})^{2N-3} (da_{2N-2}) = \tilde{c}\; (\det a_{2N-2})^{N-3/2} (da_{2N-2}).
\end{align}
Multiplying both sides of (\ref{vacovs}) by $e^{-\Tr (v_{2N-2}^{\dagger} v_{2N-2})}$ and integrating we have for the LHS
\begin{align}
\nonumber &\int e^{-\Tr (v_{2N-2}^{\dagger} v_{2N-2})} (dv_{2N-2}) = \\
\nonumber & \int_{(-\infty, \infty)^{4N-4}} e^{-2 \sum_{j=1}^{N-1}\left( |s_{j, N}|^2 + |r_{j, N}|^2 \right) } \prod_{j=1}^{N-1} d\mathrm{Re}(s_{j, N}) d\mathrm{Im}(s_{j, N}) d\mathrm{Re}(r_{j, N}) d\mathrm{Im}(r_{j, N})\\
\nonumber &=\prod_{j=1}^{4N-4} \int_{-\infty}^{\infty} e^{-2x^2} dx = \left( \frac{\pi}{2}\right)^{2N- 2},
\end{align}
and for the RHS
\begin{align}
\nonumber \tilde{c} \int e^{-\Tr \; a_{2N-2}} (\det a_{2N-2})^{N-3/2} (da_{2N-2})= \tilde{c} \int_0^{\infty} e^{-2x} x^{2N-3} dx = \tilde{c} \; \frac{\Gamma (2N-2)}{4^{N-1}}.
\end{align}
because $a_{2N-2} =\diag (x, x)$, where $x\in(0, \infty)$. We deduce that
\begin{align}
\nonumber \tilde{c} = \frac{\pi^{2N-2}}{\Gamma (2N-2)}.
\end{align}

Now to calculate (\ref{itint1})
\begin{align}
\nonumber &\int (du_{2N-2}) \det \left(\1_2 + v_{2N-2}^{\dagger} v_{2N-2} \right)^{-2 N}\\
\nonumber &= \det (\1_{2N-2} +\bR_{2N-2} \bR_{2N-2}^{\dagger}) \det (\1_2 +z_N z_N^{\dagger})^{N- 1} \int (dv_{2N-2}) \det \left(\1_2 + v_{2N-2}^{\dagger} v_{2N-2} \right)^{-2 N}\\
\nonumber &=K^{(1)}_{2N-2} \; \frac{\pi^{2N-2}}{\Gamma (2N-2)} \int \det (a_{2N-2})^{N-3/2} \det (1+a_{2N-2})^{-2N} (da_{2N-2})\\
\nonumber &=K^{(2)}_{2N-2} \int_{0}^{\infty} x^{2N-3} (1+x)^{-4N} dx = K^{(2)}_{2N-2} \int_{0}^{\infty} \left(\frac{x}{1+x}\right)^{2N- 3} \frac{1}{(1+x)^{2N+3}} \; dx,
\end{align}
where $K^{(1)}_{2N-2}$ stands for the determinants in front of the integral after the first equality, and $K^{(2)}_{2N-2}= K^{(1)}_{2N-2} \times \pi^{2N-2}/\Gamma(2N-2)$. Letting $y=x/(1+x)$, and so $dx=(1+x)^2 dy$, we find a Selberg integral \cite{Selb1944} resulting in
\begin{align}
\nonumber &K^{(2)}_{2N-2} \int_{0}^{\infty} \left(\frac{x}{1+x}\right)^{2N- 3} \frac{1}{(1+x)^{2N+3}} \; dx = K^{(2)}_{2N-2} \int_{0}^1 y^{2N-3} (1-y)^{2N+1} dy\\
\nonumber &= K^{(2)}_{2N-2} \frac{\Gamma(2N-2) \Gamma(2N+2)}{\Gamma(4N)}\\
\nonumber &= \pi^{2N-2} \frac{\Gamma(2N+2)}{\Gamma(4N)} \det (\1_{2N-2} +\bR_{2N-2} \bR_{2N-2}^{\dagger})  \det (\1_2 +z_N z_N^{\dagger})^{N-1}.
\end{align}
Substituting this into (\ref{P4split}) we have
\begin{align}
\nonumber &\int^{*} \mathcal{P}_{4}(\bY) (d\bY)= \frac{\pi^{-2(N(N-1)+1)}}{2^N \Gamma(N+1)} \frac{\Gamma(2N+2)}{\Gamma(4N)} \prod_{j=0}^{N-1} \frac{\Gamma \left( 2(N+1+j) \right)} {\Gamma \left( 2(j+1) \right)} \frac{|\tilde{\Delta} (\vec{\lambda})|^2}{(1+|\lambda_N|^2)^{2(N+ 1)}}\\
\nonumber &\times \det \left( \1_{2N-2} +\bR_{2N-2}\bR_{2N-2}^{\dagger} \right)^{-(2 N-1)} \prod_{j=1}^{N}d\lambda_j \bigwedge_{s=2}^{N-1} (du_{2N-2s})\; (\bQ^{\dagger}d\bQ),
\end{align}
where the $*$ on the integral indicates that we have only done some of the integrals required. We then iterate this procedure over each of the $N-2$ remaining columns.

Lastly, we need the integral over $(\bQ^{\dagger}d\bQ)$. As mentioned above, $\bQ\in Sp(N)/(U(1))^N$, and using a result from \cite{Nach1965} (see \cite[Ch. 7.1]{Mays2011} for more details on this point) we have
\begin{align}
\nonumber \int (\bQ^{\dagger}d\bQ) = \frac{\mathrm{Vol} (Sp(2N))} {(\mathrm{Vol} (U(1)))^N}= \frac{\pi^{N^2}}{\prod_{j=1}^N \Gamma(2j)},
\end{align}
where we have used the result from \cite[Eq. (108)]{Dyso1962a}
\begin{align}
\nonumber \mathrm{Vol} \Big(Sp(N) \Big)= \frac{2^N \pi^{N(N +1)}}{\prod_{j=1}^N \Gamma(2j)}.
\end{align}

So having integrated out all independent variables in (\ref{P4split}), except for those corresponding to the eigenvalues $\lambda_j =x_j +iy_j$, we are left with the eigenvalue jpdf
\begin{align}
\nonumber &\mathcal{Q}(\vec{\lambda}) (d\vec{\lambda}) = \frac{\pi^{-N}}{\Gamma(N+1)} \prod_{j=1}^N {\frac{\Gamma(2N+2)} {(\Gamma(2j))^2}}\\
\label{b4ejpdf} &\times \prod_{j=1}^N \frac{|\lambda_j -\overline{\lambda}_j|^2} {(1+|\lambda_j|^2)^{2(N+1)}} \prod_{1\leq j < k \leq N} |\lambda_k- \lambda_j|^2 |\lambda_k- \overline{\lambda}_j|^2 \prod_{j=1}^{N} dx_j dy_j,
\end{align}
where, for later convenience, we restrict $\lambda$ to the upper half plane (in other words we order the complex conjugate pairs $\mathrm{Im}(\lambda)> \mathrm{Im}(\overline{\lambda})$) multiplying through by $2^N$.

\subsection{Fractional linear transformation}

As in \cite{ForrMays2011} we can take advantage of the expected rotational symmetry by applying the fractional linear transformation (\ref{flt}). Then with $\lambda_j =x_j +iy_j$ and $w_j= a_j +ib_j$ we have
\begin{align}
\nonumber y_j =\frac{1-|w_j|^2} {|1+w_j|^2}&& \mathrm{and} &&dx_j dy_j= \frac{4}{|1+w_j|^4} da_j db_j,
\end{align}
in which case
\begin{align}
\nonumber &\nonumber \prod_{1\leq j < k \leq N} |\lambda_k- \lambda_j|^2 |\lambda_k- \overline{\lambda}_j|^2 = \prod_{1\leq j < k \leq N} \frac{16}{|w_j+1|^4 |w_k+1|^4} |w_k-w_j|^2 |\overline{w}_j w_k-1|^2\\
\nonumber &= \prod_{j=1}^N \left(\frac{2 \; \overline{w}_j} {|w_j+1|^2}\right)^{2(N-1)} \prod_{1\leq j < k \leq N}  (w_k - w_j) \left( \frac{1} {\overline{w}_j} - \frac{1}{\overline{w}_k} \right) \left( w_k - \frac{1}{\overline{w}_j }\right) \left(w_j- \frac{1}{\overline{w}_k} \right).
\end{align}
Also
\begin{align}
\nonumber (\lambda_j-\overline{\lambda}_j)= \frac{2}{i} \frac{\overline{w}_j} {|w_j+ 1|^2} \left(w_j - \frac{1}{\overline{w}_j}\right).
\end{align}
Substituting into (\ref{b4ejpdf}) and simplifying we have
\begin{align}
\nonumber \mathcal{Q}(\mathbf{w}) (d\mathbf{w})&= (-1)^{N( N-1)/2} \frac{(\pi)^{-N}}{\Gamma(N+1)} \prod_{j=1}^N {\frac{\Gamma(2N+2)} {(\Gamma(2j))^2}} \\
\label{ejpdf1} &\times \prod_{j=1}^N \frac{1} {|w_j|^2} \left( \frac{\overline{w}_j}{w_j} \right)^{N -1/2} \frac{\frac{1}{|w_j|}- |w_j|} {\left(|w_j|+ \frac{1}{|w_j|}\right)^{2N+2}} \; \Delta \left( \mathbf{w}, \mathbf{\frac{1}{\overline{w}}}\right) (d\mathbf{w}),
\end{align}
from which (\ref{transejpdf}) follows, where $\Delta(\mathbf{x})$ is the standard Vandermonde product. (Note that the factor of $(i)^{-1}$ in (\ref{transejpdf}) is an artifact of splitting the product $1/|w_j|- |w_j|$ into the factors of $\tau(w_j)$ and $\tau(1/\overline{w}_j)$. It was for this splitting that we ordered the complex conjugate pairs in (\ref{b4ejpdf}).)

\section{Generalized partition function}
\label{s:gpf}

We will now find a Pfaffian expression for the generalized partition function
\begin{align}
\label{e:gpf} Z_N[v]:=\Big\langle \prod_{j=1}^N v(w_j) \Big\rangle_{\mathcal{Q}} = \int_{\mathbb{D}} v(w_1)\; (dw_{1}) \cdot\cdot\cdot \int_{\mathbb{D}} v(w_N)\; (dw_{N}) \: \mathcal{Q}(\bw),
\end{align}
where $\mathbb{D}$ is the unit disk, with which we shall calculate the correlation functions in Section \ref{s:correlns}. Using the Vandermonde identity we have
\begin{align}
\nonumber \Delta\left( \bw, \frac{1}{\overline{\bw}} \right)&= \det \left[
\begin{array}{c}
\left[ p_{k-1} (w_j)  \right]_{j=1,..., N}\\
\left[ p_{k-1} (\frac{1}{\overline{w}_j}) \right]_{j=1,..., N}
\end{array} \right]_{k= 1,..., 2N}\\
\nonumber &= (-1)^{N(N-1)/2} \det \left[
\begin{array}{c}
p_{k-1} (w_j)\\
p_{k-1} (\frac{1}{\overline{w}_j})
\end{array} \right]_{\genfrac {}{}{0pt}{}{j=1,..., N} {k=1,..., 2N}}
\end{align}
and so (\ref{transejpdf}) becomes
\begin{align}
\label{ejpdfdet} \mathcal{Q}(\mathbf{w}) = (-1)^{N(N-1)/2} C_N \prod_{j=1}^N \frac{1}{i \: |w_j|^2} \tau(w_j) \tau \left(\frac{1} {\overline{w}_j} \right) \det \left[
\begin{array}{c}
p_{k-1} (w_j)\\
p_{k-1} (\frac{1}{\overline{w}_j})
\end{array} \right]_{\genfrac {}{}{0pt}{}{j=1,..., N} {k=1,..., 2N}}.
\end{align}
Before proceeding we will reorder the columns of the determinant in (\ref{ejpdfdet}), as in \cite[eqn. (30)]{ForrMays2011}, as follows
\begin{align}
\label{e:pord}
p_0,p_{N-1},p_2,p_{N-3},\cdot\cdot\cdot , p_{N-2},p_1,
\end{align}
introducing another factor of $(-1)^{(N/2)(N/2-1)/2}$. For labeling purposes we define
\begin{align}
\nonumber q_{2j}(x)=p_{2j}(x),\qquad q_{2j+1}(x)=p_{N-1-2j}(x).
\end{align}
This reordering will make finding the relevant skew-orthogonal polynomials straightforward (see Section \ref{s:sops}). With the relabeling we have
\begin{align}
\label{ejpdfdet1} \mathcal{Q}(\mathbf{w}) = C_N \prod_{j=1}^N \frac{1}{i \: |w_j|^2} \tau(w_j) \tau \left(\frac{1} {\overline{w}_j} \right) \det \left[
\begin{array}{c}
q_{k-1} (w_j)\\
q_{k-1} (\frac{1}{\overline{w}_j})
\end{array} \right]_{\genfrac {}{}{0pt}{}{j=1,..., N} {k=1,..., 2N}}.
\end{align}

Substituting this into (\ref{e:gpf}), expanding out the determinant using the Leibniz formula, and then ordering the terms to give a Pfaffian we obtain the following result (see \cite[Chapter 6.1.2]{Forr2010} for explicit details).
\begin{proposition}
The generalized partition function for the real quaternion spherical ensemble can be written as
\begin{align}
\label{gpf1} Z_{N} [v] &= \Gamma [N+1] \; C_N\Pf [\gamma_{j,k} [v]]_{j,k= 1,...,2N},
\end{align}
where
\begin{align}
\nonumber \gamma_{j,k}[v] := \frac{1}{i} \int_{\mathbb{D}} \frac{v(w)}{|w|^2} \tau(w) \tau \left(\frac{1} {\overline{w}} \right) \left(q_{j-1}(w) q_{k-1}\left(\frac{1} {\overline{w}} \right) - q_{j-1}\left(\frac{1} {\overline{w}} \right) q_{k-1}(w) \right) (dw).
\end{align}
\end{proposition}

\section{Skew-orthogonal polynomials}
\label{s:sops}

In this section we find the monic polynomials that will reduce the Pfaffian in (\ref{gpf1}) (with $v=1$) to block diagonal form, where the diagonal blocks are the $2\times 2$ matrices
\begin{align}
\nonumber \left[ \begin{array}{cc}
0 & h_j\\
-h_j & 0
\end{array}\right],
\end{align}
in which case
\begin{align}
\label{e:ZN[1]} Z_N[1]= \Gamma [N+1] \; C_N \prod_{j=1}^N h_j.
\end{align}
In other words, define the skew-symmetric inner product
\begin{align}
\nonumber \langle q_j, q_k \rangle := \gamma_{j+1,k+1}[1]
\end{align}
and look for polynomials to satisfy the skew-orthogonality conditions
\begin{align}
\label{e:soc} \langle p_{2j},p_{2k}\rangle = \langle p_{2j+1}, p_{2k+1} \rangle=0 &, &\langle p_{2j}, p_{2k+1} \rangle= -\langle p_{2k+1} ,p_{2j} \rangle= \delta_{j,k} \: h_j.
\end{align}

By writing out $\gamma_{j,k}[1]$ explicitly, we see that the only time it is non-zero is when the angular dependence cancels. With this fact, and the convenient reordering in (\ref{e:pord}), it can be seen that the skew-orthogonal polynomials are the monomials
\begin{align}
\label{sops} q_{2j}(w)= w^{2j} ,&& q_{2j+1}(w)= w^{2N-1-2j}.
\end{align}
With $w=r e^{i\theta}$ we substitute these polynomials into $\gamma_{j,k}[1]$ to obtain
\begin{align}
\nonumber h_j&= 2\pi \int_0^1 \frac{1-r^2}{(1+r^2)^{2N+2}} \left( r^{4j+1}- r^{4N-4j-1} \right)\; dr.
\end{align}
By changing variables $c= r^2$ we can make use of the identity in \cite[Equation 3.216 (1)]{GradRyzh2000}: for general $a,b$ such that Re$\, b>0$,
Re$\, (a-b) > 0$,
\begin{align}
\nonumber \int_0^1(t^{b-1} + t^{a-b-1})(1 + t)^{-a} \, dt = \frac{\Gamma(b)\Gamma(a-b)}{\Gamma(a)},
\end{align}
and so, with $b = y$, $a-b = x$, we have a non-standard form of the beta integral
\begin{align}
\nonumber \int_0^1 t^{x-1}(1-t)^{y-1} \, dt = \frac{\Gamma(x)\Gamma(y)}{\Gamma(x+y)}.
\end{align}
This gives
\begin{align}
\nonumber h_j&= \frac{\pi (2N-4j-1)}{(2N+1) (2N)} {2N-1 \choose 2j}^{-1},&& j=0,..., N-1,
\end{align}
and, from (\ref{gpf1}),
\begin{align}
\label{ZN[1]} \prod_{j=0}^{N-1} h_j = \frac{1}{C_N \Gamma(N+1)}.
\end{align}

\section{Correlation functions}
\label{s:correlns}

To calculate the $n$-point correlation functions we will use the functional differentiation formula
\begin{align}
\label{ZS1}
\rho_{(n)}(\mathbf{w})
= \frac{1}{Z_N [v]} \frac{\delta^{n}} {\delta v(w_1) \cdots \delta v(w_{n})} Z_N [v] \Big |_{v=1}.
\end{align}
Indeed, by the skew-orthogonality (\ref{e:soc}), it is straightforward to calculate the $1$-pt correlation function (the eigenvalue density) by applying (\ref{ZS1}) to (\ref{gpf1}),
\begin{align}
\nonumber \rho_{(1)}(w) &= \frac{1}{Z_N[1]} \left( \sum_{j=1}^{N} \left( \frac{\partial} {\partial v(w)} \gamma_{2j-1, 2j}[v] \right) \prod_{\genfrac{}{}{0pt}{}{k=0}{k\neq j-1}}^{N-1} h_k \right),
\end{align}
which reduces to (\ref{sxx1}) by making use of (\ref{e:ZN[1]}) and (\ref{ZN[1]}).

For the general correlations, we make use of the formula $\det (\1 +\bA\bB)= \det (\1 + \bB\bA)$, or its quaternion determinant analogue $\mathrm{qdet} (\1 +\bA\bB)= \mathrm{qdet}(\1 + \bB\bA)$ (for $\bB\bA$ self-dual) --- or an equivalent Pfaffian formula, see \cite{Rain2000} --- in conjunction with the Fredholm quaternion determinant
\begin{align}
\nonumber \qdet[1+\lambda K]:=1+\sum_{s=1}^{\infty}\frac{\lambda^s}{s!}\int_{-\infty}^{\infty}dx_1\cdot\cdot\cdot\int_{-\infty}^{\infty}dx_s\;\qdet [K(x_j,x_k)]_{j,k=1,...,s},
\end{align}
where $K$ is an integral operator with $2\times 2$ kernel $K(x, y)$, and the matrix $[K(x_j,x_k)]_{j,k=1,...,s}$ is self-dual. The details are lengthy and tedious, however it is directly analogous to the same approach for the real spherical and real Ginibre ensembles, see \cite{Mays2011} for a detailed exposition (the techniques therein were inspired by those in \cite{BoroSinc2009} and \cite{Forr2010}). We find that the correlation functions are of the same structure as those for the real spherical ensemble \cite{ForrMays2011}.

\begin{proposition}
\label{thm:correlns}
With the eigenvalues transformed according to (\ref{flt}), then using the skew-orthogonal polynomials (\ref{sops}) the $n$-point eigenvalue correlation functions for the real quaternion spherical ensemble are
\begin{align}
\label{correlns} \rho_{(n)}(w_1,..., w_{n})=\mathrm{Pf}\left[
K_N(w_l,w_m)\right],\qquad w_i \in \mathbb{D},
\end{align}
\begin{align}
\label{kernel} K_N(x, y)=\left[\begin{array}{cc}
D(x, y) & S(x, y)\\
-S(y, x) & I(x, y)\\
\end{array}\right],
\end{align}
where 
\begin{align}
\nonumber D(x, y)&= \frac{1}{i} \sum_{j=0}^{N -1}\frac{1}{h_j}\Big( a_{2j}(x) a_{2j+1}(y)-a_{2j+1}(x) a_{2j}(y) \Big),\\
\nonumber S(x, y)&= \frac{1}{i} \sum_{j=0}^{N -1}\frac{1}{h_j}\Big( a_{2j}(x) b_{2j+1}(y)-a_{2j+1}(x) b_{2j}(y) \Big),\\
\nonumber I(x, y)&= \frac{1}{i} \sum_{j=0}^{N -1}\frac{1}{h_j}\Big( b_{2j}(x) b_{2j+1}(y)-b_{2j+1}(x) b_{2j}(y) \Big),
\end{align}
and
\begin{align}
\nonumber a_j(x) &= |x|^{-1} \tau(x) \hspace{2pt}q_j(x),\\
\nonumber b_j(x) &= |x|^{-1} \tau \left(\frac{1} {\overline{x}}\right) \hspace{2pt}q_j \left(\frac{1} {\overline{x}} \right).
\end{align}
\end{proposition}

Substituting in the various factors we have
\begin{align}
\nonumber &S(w, z)= A_{wz} \frac{(2N+1) (2N)} {\pi} \sum_{j=0}^{N-1} {2N-1 \choose 2j} \frac{\left( (w \bar{z})^{2j+1/2 -N} - (w \bar{z})^{N-1/2-2j} \right)} {2N-4j-1}\\
\label{sxy1} &= A_{wz} \frac{(2N+1) (2N)} {\pi} \sum_{j=0}^{N-1} {2N-1 \choose j} \frac{\left( (w \bar{z})^{j+1/2 -N} - (w \bar{z})^{N-1/2-j} \right)} {2N-2j-1},
\end{align}
where
\begin{align}
\nonumber A_{wz}= \frac{1} {r_w r_z} \frac{\left( \frac{1}{r_w}- r_w \right)^{1/2} \left( \frac{1}{r_z}- r_z \right)^{1/2}} {\left( r_w + \frac{1} {r_w} \right)^{N+1} \left( r_z + \frac{1} {r_z} \right)^{N+1}},
\end{align}
with $|w|=r_w$ and $|z|= r_z$. The other kernel elements satisfy the relations
\begin{align}
\nonumber I(w, z) &=\frac{1}{|w|^2} \; S\left( \frac{1} {\overline{w}}, z \right),\\
\label{e:ISD} D(w,z) &=\frac{1}{|w|^2} \; S\left( w, \frac{1} {\overline{z}} \right).
\end{align}

In the following, we analyze the eigenvalue density, which includes several checks of our working, however we first rewrite $S(w, z)$. Let
\begin{align}
\nonumber \sigma(\gamma):= \sum_{j=0}^{N-1} {2N-1 \choose j} \frac{\left( \gamma^{j+1/2 -N} - \gamma^{N-1/2-j} \right)} {2N-2j-1},
\end{align}
then
\begin{align}
\nonumber \gamma \frac{d \sigma(\gamma)}{d\gamma} = - \frac{(1+\gamma)^{2N-1}}{2 \gamma^{N-1/2}},
\end{align}
by anti-differentiating we have 
\begin{align}
\nonumber \sigma(1)- \sigma(v) = \int_v^1 \frac{d \sigma(\gamma)} {d\gamma} d\gamma= -\frac{1}{2} \int_v^1 \frac{(1+\gamma)^{2N-1}} {\gamma^{N+1/2}} d\gamma.
\end{align}
From the definition of $\sigma(\gamma)$ we see that $\sigma(1)=0$ and therefore
\begin{align}
\nonumber \sigma(v)&= 2^{2N-2} \int_{0}^{1-v} (1-\gamma)^{-N-1/2} (1-\gamma/2)^{2N-1} d\gamma,
\end{align}
where we have changed variables $\gamma \mapsto 1-\gamma$, thus giving the integral expression
\begin{align}
\label{e:swz1} S(w, z)= A_{wz} \frac{2^{2N-1} (2N+1) N}{\pi} \int_{0}^{1- w\overline{z}} (1-\gamma)^{-N-1/2} (1-\gamma/2)^{2N-1} d\gamma,
\end{align}
which we will make use of in analyzing the asymptotic behaviour of the correlation functions.

\subsection{Eigenvalue density}
\label{s:dens}

\begin{figure}[htp]
\begin{center}
\includegraphics{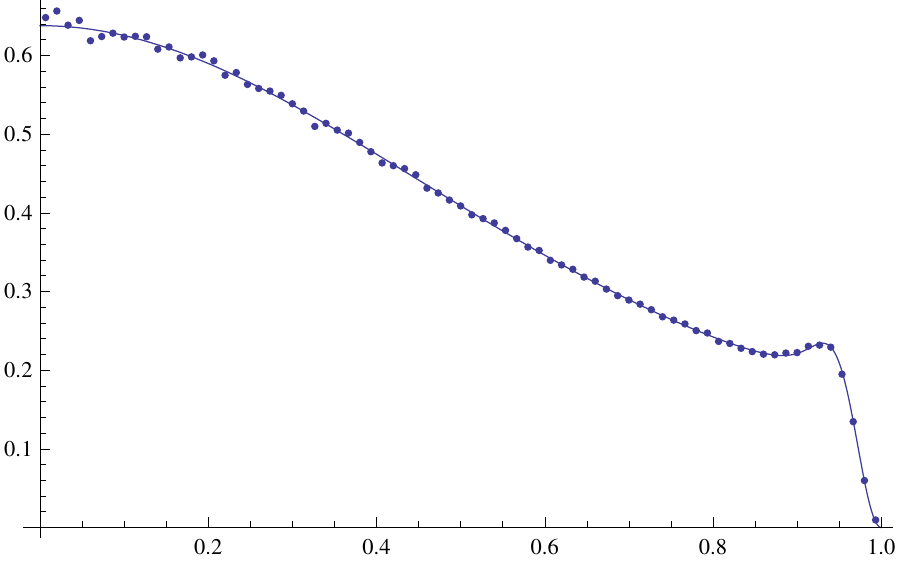}
\end{center}
\caption{The points are the bin mid-points for a simulated radial (normalized) density of eigenvalues of $2000$ instances of $500\times 500$ matrices, where the eigenvalues have been transformed according to (\ref{flt}). The noisiness of the data near $r=0$ is due to the relatively small number of eigenvalues near the origin, which is a consequence of the relatively smaller area one is sampling from. The solid line is a plot of (\ref{sxx1}), scaled by $1/N$.}
\label{f:edens}
\end{figure}

With $n=1$ (\ref{correlns}) reduces to $\rho_{(1)} (w)= S(w, w)$. We see from (\ref{sxy1}) and (\ref{e:swz1}) that this immediately gives us the two expressions for the density in (\ref{sxx1}). We can check that
\begin{align}
\nonumber \int_0^{2\pi}d\theta \int_0^1 r dr \; \rho_{(1)} (r e^{i \theta}) = N
\end{align}
as required. For another check, in Figure \ref{f:edens} we find good agreement between (\ref{sxx1}) and a simulated eigenvalue density. Note the elevated density near $r=1$, which we recall corresponds to the real line. We interpret this as the effect of repulsion from the real line, as discussed in the Introduction.

To analyze the behaviour for large $N$ we first exponentiate the logarithm of the integrand in (\ref{sxx1}) giving
\begin{align}
\nonumber \int_{0}^{1- r^2} \exp \left[ (2N-1) \log (1-\gamma/2) - (N+1/2) \log (1-\gamma) \right] d\gamma.
\end{align}
For large $N$ this integral will be dominated by points close to the upper terminal, so we Taylor expand
the exponent about $1-r^2$
\begin{align}
\nonumber &(2N-1) \log (1-\gamma/2) - (N+1/2) \log (1-\gamma)\\
\nonumber & \approx \log \left( \frac{(1+r^2)^{2N-1}} {2^{2N-1} r^{2N+1}} \right)+ (\gamma -1 +r^2) \left( \frac{2N+1}{2r^2} - \frac{2N-1}{1+r^2} \right)
\end{align}
and so for large $N$ the integral in (\ref{sxx1}) is approximated by
\begin{align}
\nonumber \frac{(1+r^2)^{2N-1}} {2^{2N-1} r^{2N+1}} \int_0^{1-r^2} \exp \left[ (\gamma -1 +r^2) \left( \frac{2N+1}{2r^2} - \frac{2N-1}{1+r^2} \right) \right] d\gamma.
\end{align}
Substituting this into (\ref{sxx1}) and taking $N$ to infinity (noting that $r^2-1<0$) yields (\ref{e:limdens}).

\subsection{Scaled limit}
\label{s:slims}

Here we aim to recover the bulk real quaternion Ginibre correlation functions of \cite{Kanz2002}, by zooming in on the origin. Under the transformation (\ref{flt}) the origin in the original coordinates is mapped to the point $(1,0)$ and so we make the replacements
\begin{align}
\nonumber w\mapsto w^{\sharp}&= 1+2i W \sqrt{\frac{\pi} {N}}, \qquad W=X+iY,\\
\label{e:slims} z\mapsto z^{\sharp}&= 1+2i Z \sqrt{\frac{\pi} {N}}, \qquad Z=A+iB,
\end{align}
in (\ref{e:swz1}) and look for the large $N$ limit of
\begin{align}
\label{swzsharp} \frac{4 \pi}{N} S(w^{\sharp}, \overline{z}^{\sharp}) = 2^{2N+1} (2N+1) A_{w^{\sharp} z^{\sharp}} \int_{0}^{1- w^{\sharp} \overline{z}^{\sharp}} (1-\gamma)^{-N-1/2} (1-\gamma/2)^{2N-1} d\gamma.
\end{align}
We have chosen the transformations (\ref{e:slims}) to make a correspondence with the results in \cite{Kanz2002}, however we note that the density will be twice as large as in that work since here we have restricted the eigenvalues to be in the upper half plane.

First rewrite $A_{w^{\sharp} z^{\sharp}}$ as
\begin{align}
\nonumber A_{w^{\sharp} z^{\sharp}}= (r_{w^{\sharp}} r_{z^{\sharp}})^{N-1/2} \frac{(1-r_{w^{\sharp}}^2)^{1/2} (1-r_{z^{\sharp}}^2)^{1/2}} {(1+r_{w^{\sharp}}^2)^{N+1} (1+r_{z^{\sharp}}^2)^{N+1}},
\end{align}
where
\begin{align}
\nonumber r_{w^{\sharp}}= \left( 1- \frac{4 \sqrt{\pi} Y} {\sqrt{N}} + 4\pi \frac{|W|^2} {N} \right)^{1/2}.
\end{align}
Using the asymptotic behaviour
\begin{align}
\label{e:xsnn} \left( 1+\frac{x}{\sqrt{N}} \right)^{N} \sim e^{x \sqrt{N}- x^2/2},
\end{align}
which holds up to terms of order $1/\sqrt{N}$ in the exponent, we find that
\begin{align}
\nonumber r_{w^{\sharp}}^{N-1/2} & \sim \exp \left( -2Y\sqrt{\pi N} +2\pi |W|^2 - 4 \pi Y^2\right)\\
\nonumber (1+r_{w^{\sharp}}^2)^{N+1}& \sim 2^{N+1} \exp \left[ 2\pi |W|^2- 2\sqrt{\pi N} \: Y - 2 \pi Y^2 \right],\\
\nonumber (1- r_{w^{\sharp}}^2)^{1/2}& \sim \left(\frac{4 \sqrt{\pi}\: Y}{\sqrt{N}} \right)^{1/2},
\end{align}
and so
\begin{align}
\nonumber A_{w^{\sharp} z^{\sharp}} \sim \frac{\sqrt{\pi} \left( YB \right)^{1/2}} {2^{2N} \sqrt{N}} \; e^{-2\pi (Y^2+ B^2)}.
\end{align}

Ignoring terms of order $1/N$ we take
\begin{align}
\nonumber v= 1+2i W \sqrt{\frac{\pi} {N}} - 2i\overline{Z} \sqrt{\frac{\pi} {N}} \approx w^{\sharp} \overline{z}^{\sharp},
\end{align}
and the integral in (\ref{swzsharp}) becomes
\begin{align}
\nonumber \int_{0}^{-2i \sqrt{\frac{\pi} {N}} (W-\overline{Z})} (1-\gamma)^{-N-1/2} (1-\gamma/2)^{2N-1} d\gamma \sim \int_{0}^{-2i \sqrt{\frac{\pi} {N}} (W-\overline{Z})} \left( 1+\frac{\gamma^2}{4} \right)^N d\gamma,
\end{align}
since, for large $N$, $\gamma$ is restricted to lie close to zero. Then, by changing variables $\gamma \mapsto -2i\gamma/\sqrt{N}$ the integral approaches
\begin{align}
\nonumber \frac{ \sqrt{\pi}} {i \sqrt{N}} \; \erf \left( \sqrt{\pi} (W-\overline{Z}) \right).
\end{align}

Combining the previous we have the large $N$ behaviour of (\ref{swzsharp}).
\begin{proposition}
\label{p:slims}
Changing to the variables $w^{\sharp}$ and $z^{\sharp}$ from (\ref{e:slims}) we find (\ref{e:swz1}) has the limiting behaviour
\begin{align}
\label{e:slswx} \frac{4 \pi}{N} S(w^{\sharp}, \overline{z}^{\sharp}) &\mathop{\sim}\limits_{N\to \infty} \frac{4 \pi} {i} \left( YB \right)^{1/2} e^{-2\pi (Y^2+ B^2)} \; \erf \Big( \sqrt{\pi} (W-\overline{Z}) \Big),
\end{align}
which, with $W=Z$, gives us the scaled limit of the density
\begin{align}
\nonumber 16 \pi Y^2 e^{-4\pi Y^2}\int_{0}^{1} e^{4\pi Y^2u^2} du.
\end{align}
\end{proposition}

The scaled limit of the remaining kernel elements ($D$ and $I$) can be obtained similarly, or from (\ref{e:slswx}) by using formulae analogous to (\ref{e:ISD}) that apply to the real quaternion Ginibre ensemble
\begin{align}
\nonumber I(w, z)= i S(\overline{w}, z),\\
\nonumber D(w, z)= -i S(w, \overline{z}).
\end{align}
The result from \cite{Kanz2002} corresponding to (\ref{e:slswx}) is
\begin{align}
\label{e:Kd} \frac{2 \pi} {i} (YB)^{1/2} e^{-\pi (|W|^2+ |Z|^2)+ \pi (\overline{Z}^2+ W^2)} \mathrm{erf} \Big( \sqrt{\pi} (W-\overline{Z}) \Big).
\end{align}
Ignoring the imaginary parts of (\ref{e:Kd}), which always cancel in Kanzieper's formulation, we can use the properties of the complex error function to see that the correlation functions for the real quaternion spherical ensemble agree with those of the corresponding Ginibre ensemble under the scaling limit implied by (\ref{e:slims}) (up to the aforementioned factor of $2$). Further, from the recent results in \cite{TaoVu2012}, we see that in this limit, we recover the universal correlation functions for planar (iid, with first four moments finite) ensembles.

\section{Discussion}

As discussed in the introduction, a tripartite categorization of matrix ensembles into geometrically-themed universality classes was suggested in \cite{Mays2011} (which was also used in a similar context in \cite{Kris2006}). Each of these classes corresponds to a surface of uniform curvature: the plane, the sphere and the anti- or pseudo-sphere. The Ginibre ensembles, being ensembles of iid matrices, obey the circular law, which states that the eigenvalues are uniformly distributed on a planar disk in the limit of large matrix dimension, and so we may associate the Ginibre ensembles with the plane.

Taking two of these iid matrices and forming the product $\bY=\bA^{-1}\bB$, which is a matrix generalization of a Cauchy random variable, gives us spherical ensembles. The spherical law (analogously to the circular law) states that the eigenvalues have uniform density on a sphere (under stereographic projection) in the limit of large matrix dimension and so these ensembles are naturally associated with the sphere. In this work we have applied the (skew-)orthogonal polynomial method to calculate the correlation functions for the real quaternion spherical ensemble, which is the $\beta=4$ analogue of the ensembles considered in \cite{Kris2006, Kris2009, HKPV2009} and \cite{ForrMays2011}. This completes the exploration of the matrices in the spherical universality class using the Dyson indices $\beta= 1,2,4$. As the spherical law was established in \cite{Bord2010} by adapting the proof of the circular law in \cite{TaoVuKris2010}, we suspect it is possible to adapt the proof in \cite{TaoVu2012} to show that the general $n$-point correlation functions for the complex spherical ensemble are universal for the spherical class of matrices.

The third universality class in this categorization corresponds to the anti-sphere, which is a surface of constant negative curvature. The ensembles in this class are given by sub-blocks (or truncations) of unitary matrices. To date, truncations of complex unitary \cite{ZyczSomm2000, PetzReff2003, Forr2006} and real orthogonal \cite{KhorSommZycz2010} matrices have been analyzed (for a discussion of the real orthogonal case also see \cite{Mays2011}). Naturally enough, the next step would seem to be an analysis of a  $\beta=4$ real quaternion analogue of truncated unitary and orthogonal ensembles. Further, to conclude the program laid down in \cite{Mays2011}, perhaps by again adapting the proof of the circular law in \cite{TaoVuKris2010}, then a proof of a so-called `anti-spherical' law may be achieved (see \cite[Chapter 7.6.2]{Mays2011} for more on this), and, even further, perhaps using the methods of \cite{TaoVu2012} find that the general correlations are universal.

\section*{Acknowledgements}

AM was partly supported by the Australian Mathematical Society (AustMS) LiftOff Fellowship during the work leading to this paper. Special thanks to Peter Forrester for reading an early draft of this work and providing many useful comments. The author would also like to thank Gernot Akemann, Tilo Wettig and Anita Ponsaing for several interesting discussions, and the following for their kind hospitality: Department of Mathematics, Vanderbilt University; Department of Mathematics, University of Geneva; Faculty of Physics, University of Bielefeld; and Faculty of Physics, University of Regensburg. Thanks to Joshua Feinberg for pointing me to an earlier work of his which is relevant to this paper.

\end{document}